\documentclass[conference]{IEEEtran}
\usepackage{amsmath, bm, amssymb, amsfonts}
\usepackage{enumerate}
\usepackage{graphicx}
\usepackage{color}
\usepackage{url}
\usepackage[noadjust]{cite}

\DeclareMathOperator*{\argmax}{\arg\!\max}

\begin{document}

\title{A Latent Social Approach to YouTube Popularity Prediction}
\author{\IEEEauthorblockN{Amandianeze O Nwana, Salman Avestimehr, and Tsuhan Chen}
\IEEEauthorblockA{School of Electrical and
Computer Engineering\\
Cornell University\\
Ithaca, New York 14853\\
Email: aon3@cornell.edu, \{avestimehr, tsuhan\}@ece.cornell.edu}
}

\maketitle

\begin{abstract}

Current works on Information Centric Networking assume the spectrum of
caching strategies under the Least Recently/Frequently Used (LRFU) scheme
as the de-facto standard, due to the ease of implementation and easier analysis of
such strategies.
In this paper we predict the popularity distribution of YouTube videos within a campus
network. We explore two broad approaches in predicting the 
popularity of videos in the network:  consensus approaches based on aggregate
behavior in the network, and social approaches based on the information 
diffusion over an \textit{implicit} network. 
We measure the performance of our approaches under a simple caching framework
by picking the \textit{k} most popular videos according to our predicted distribution
and calculating the hit rate on the cache.
We develop our approach by first incorporating video inter-arrival
time (based on the power-law distribution governing the transmission time between 
two receivers of the same message in scale-free networks) to the baseline (LRFU), then combining with 
an information diffusion model over the inferred \textit{latent} social graph that governs
diffusion of videos in the network.\\
We apply techniques from latent social network inference
to learn the sharing probabilities between users in the network and apply
a virus propagation model borrowed from mathematical epidemiology
to estimate the number of times a video will be accessed in the future. Our approach 
gives rise to a 14\% hit rate improvement over the baseline.

\end{abstract}

\begin{figure}[!h]
	\centering
	\includegraphics[width=3.5in]{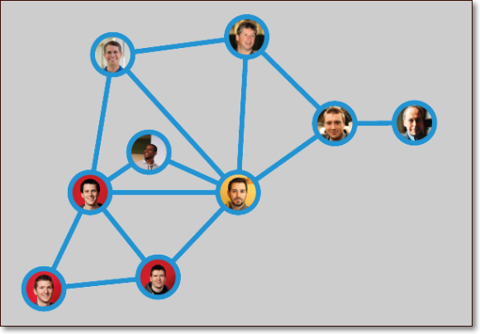}
	\caption{This social graph could represent a network of 9 users, where the edges between the users
					represent the probability they share YouTube videos with each other.}
	\label{social_graph}
	\end{figure}

\begin{section}{Introduction}

In large networks, we face the problem of having many users share the limited network resources
while expecting a certain level of quality guarantee. There are many
ways of addressing this issue like limiting the number of users that can access the network,
scaling out the network equipment (which is expensive), or moving the content closer to the
users and storing on cheap proxy caches so as to reduce end-to-end network activity.\\
In this work we focus on the caching perspective to this problem. The idea behind caching is
that many of the requests made by the users of the network are for the same objects, so
to minimize the end-to-end delay of requests in the network, the local
network should store the items that are likely to be requested again, thereby eliminating
the round trip delay that would have been experienced by these requests and simultaneously
freeing up bandwidth and other network resources.\\
We demonstrate the gains that implementing cache policies while considering the social networking
nature of video-on-demand requests like YouTube can give over the standard LRFU spectrum of
policies.
The LRFU approaches predict the popularity distribution by computing the Combined Recency and
Frequency (CRF) value for each object and caching according to the scores. Computing the CRF is
dependent on the weighing parameter $\gamma$, which takes on values from 0 to 1 and controls 
the tradeoff between recency and frequency\cite{LeeLrfu:01, LiApprox:12}. The two extremes degenerate to the Least Recently Used (LRU)
when $\gamma$ is 1, and Least Frequently Used (LFU) when $\gamma$ is 0.\\
{\bf Contributions:} We believe we are the first to propose predicting the popularity distribution
by taking into account the diffusion of these videos in the network. We incorporate diffusion in two ways, first by using the inter-arrival time
between requests for the same video for prediction, and then by using a virus propagation model
over the \textit{latent} social graph to model the spread of the videos between users. 
Because our approach is trace-driven, this social graph is latent. That is, we don't have explicit friendship information
as found on sites such as Facebook. Instead, we infer the (directed) edge weights between users in the graph, which are interpreted as the
transmission/sharing probability between two users. An example of a social graph is seen in \figurename\ref{social_graph}.\\
\end{section}

\begin{section}{Related Works}

Caching is the natural framework by which we analyze and measure the effectiveness of our approaches
in predicting the popularity of YouTube videos within the network versus other approaches in predicting 
the popularity distribution. We do this because other works such as \cite{Zink:09, Wang:12,Gill:07},
have explored the usefulness of caching YouTube and other online video requests and shown that there
are gains to be had by using a cache. \\
There are several differences between our work and these. In the work by
Wang et al. \cite{Wang:12}, they observe and track actual friendships within social networks (Facebook, Twitter, etc) and use the 
patterns of propagation for the use of replicating the content in different sites, whereas in our work, we do
not have access to the actual friendships derived from such social networks but instead 
infer unknown relationships (friendships) for predicting the 
popularity distribution of videos in the network under observation. In other works \cite{Zink:09, Gill:07},
they show that the global popularity distribution from the content provider (YouTube.com) is highly uncorrelated
with local popularity on local networks.\\
In recent years, there has been a growing interest in the field of social network analysis and
its applications in real-world computational problems. A social network can be described
as any network where the realized flow of objects over the links and nodes that
make up the network is driven by human action or behavior. Examples include road networks, recommendation networks. \\
In some situations, the human actions are directly observed on application-level
networks like Facebook, Twitter, and other social-media websites, where the links between the users are 
explicit. There are also many situations where the human-driven spread of objects in the network
is not directly observed over the links. In such cases, to understand the relationship between 
the users, we must be able to infer from the observed network transactions, 
the links between these users.
One example is the network between a city population and the spread of a virus over the population. In this case,
the spread of an infectious virus over the population are the hidden transactions, while the observable transactions are the various records
of infections by clinics, or pharmaceutical drug sales. There have been several works that have addressed the issue of latent social 
network (link) inference \cite{Liben-Nowell:03,DeChoudhury:10, GRodriguez:10, Myers:10}. 

\end{section}

\begin{section}{Approach}
	\label{p_state}
	In our problem, at a given time $t \in \mathbb{R}$, we observe some user $u \in \mathcal{U}(t,w)\subseteq \mathbb{N}$ 
	making a request, where $\mathcal{U}(t,w)$ is the	set of users at time $t$ that made requests in the past $w$ time units. 
	The users are uniquely identified by the order in which they first made requests in the network.
	Similarly at some time $t$, we observe a video $v \in \mathcal{V}(t,w) \subseteq \mathbb{N}$  being requested, 
	where $\mathcal{V}(t,w)$ is the set of videos at time $t$ that were requested in the past $w$ time
	units. The videos are also uniquely identified by the order in which they were first requested.
	In the rest of this paper, for convenience of notation we represent these	sets as $\mathcal{U}$ and $\mathcal{V}$. 
	We represent these transactions between users and videos as a triplet $(u,v,t)$, and the set of all such triplets in
	a given interval as the network trace, $\mathcal{T}(t,w)$.
	Similar to the Single Cache Performance Approximation in \cite{LiApprox:12}, our goal is to predict for some time $t$, in the 
	future, the popularity distribution of these videos given a history of these triplets before time $t$.
	In predicting the popularity distribution, we explore two general approaches, consensus-based approaches
	dependent on aggregate information in the network, and social-based approaches dependent on explicit information
	diffusion over nodes in the network. These approaches are evaluated under the framework of caching the $k$ most popular videos.\\
	Given the network trace $\mathcal{T}$,  we can calculate $\hat{X}(t,k)$ as the $k$-sparse binary vector of length $|\mathcal{V}|$,
	representing the $k$ videos to be cached at time $t$, and	$X(t) \in \mathbb{N}^{|\mathcal{V}|}$, 
	the vector representing the number of views each video got at time $t$. The hit rate, $H(\hat{X}(t,k)|\mathcal{T}(t,w))$, is then given
	as follows,
	 \begin{equation}
	 	H(\hat{X}(t,k)|\mathcal{T}) = \dfrac{X(t) \cdot \hat{X}(t,k)}{|| X(t)||_1}
	\end{equation}
	For the purpose of this paper we don't allow partial caching of videos, and we assume all videos
	have the same size.
	Caching schemes can be divided into two steps. First, the assignment of scores to
	the various objects or candidates for the cache. Second, an ordering of these scores
	to decide which objects will end up in the cache. It is not necessary that these two
	steps are done separately. Some caching strategies, for example the commonly used Least 
	Recently Used (LRU) scheme, implicitly combine the score assignment and the ordering.\\
	We discuss the approaches we explored in the following sections.

	\begin{subsection}{Baseline}
		\label{approach:baseline}
		The de-facto standard for our application is caching according to the CRF values assigned
		to each video as shown in \cite{LiApprox:12}. In \cite{LiApprox:12}, they come up with 
		with a model to approximate the performance (hit rate) of the popularity distribution
		under the LRFU assumption independent of the actual order the videos arrive, and they
		demonstrate through trace-based simulations that the approximate model is a good 
		approximate of the actual trace-based simulations	using the LRFU scheme (within	5\%). 
		We compare our approaches to the approximate popularity distribution.
	\end{subsection}

	\begin{subsection}{Consensus Approach}
		Any approach to caching videos from network traces, $\mathcal{T}$, that ignores 
		the actual users/watchers of the videos is a consensus approach.
		Such methods rely on aggregate or average properties derived from the video
		request patterns, and not particularly how a video diffuses over
		the nodes in the network. This section discusses some consensus approaches we explored.\\
		In improving the baseline, we will explore its deficiencies. A deficiency of the the baseline 
		is its handling of time and recency. Since underlying the baseline distribution is the LRFU
		scheme, it is known that the distribution is susceptible to object staleness as seen pointed out in \cite{LeeLrfu:01,Lru-k:93}.
		This is because objects can still have high CRF scores under LRFU even after a long time has elapsed since its last 
		request.
		We will consider two (orthogonal) ways of incorporating time or the notion of staleness into the 
		baseline.\\
		\underline{\bf{Viralness.}} A deficiency of the baseline that we try to capture is the expected growth 
		of a given video. Viralness tries to measure the growing popularity that a video might be experiencing.
		Let $\mathcal{D}_v(f,t,w)$ be the set of the first $f \cdot 100\%$ of transactions on video $v$, that
		occurred over the time interval $[t-w,\ t)$, with $w$ as the window size of history. We call this
		set the \textit{cascade} of video $v$.\\
		We define,
		\begin{center}
			$\lambda(v,f_1,f_2,t,w)$ - The time from the $f_1^{th}$ percentile of views to the 
							$f_2^{th}$ percentile of video $v$ on the interval $[t-w,\ t)$\\
			\ \\
			$l_v(f,t,w)$ - The time of the $f^{th}$ percentile transaction with video $v$ on interval $[t-w\ t)$\\
			\ \\
			$\lambda(v,f_1,f_2,t,w) = l_v(f_2,t,w) - l_v(f_1,t,w)$\\
			\ \\ 
			$W(v,f_1,f_2,t,w)\ - \ The\  \# \  of\  views\  in\  \lambda(v,f_1,f_2,t,w)$\\
			\ \\
			$W(v,f_1,f_2,t,w) = |\mathcal{D}_v(f_2,t,w)| - |\mathcal{D}_v(f_1,t,w)|$\\
			\ \\
			$R(v,f_1,f_2,t,w) = \frac{W(v,f_1,f_2,t,w)}{\lambda(v,f_1,f_2,t,w)}$ \\
			\ \\
			$\rho(v,f_1,f_2,f_3,t,w) = \frac{R(v,f_1,f_3,t,w)}{R(v,f_1,f_2,t,w)}$\\
		\end{center}
		The score is then given by,
		\begin{equation}
			S_{\text{viral}}(v,f_1,f_2,f_3,t,w) = \rho\cdot  R(v,f_1,f_3,t,w)
		\end{equation}
		Intuitively, this score is the number of views per video per unit time, normalized 
		by its growth trend, $\rho$. If $\rho \geq 1$. Then we are more confident that the number of
		views per unit time is indeed growing. This is akin to a second derivative test, while the rate of
		views is akin to the first derivative. The ordering here is also a sort in decreasing order of viral 
		score.
		\ \\
		\underline{\bf{Inter-arrival Time.}} In this approach we model the average 
		inter-arrival/inter-viewing times of videos in the network.
		To do this, for each video we take the average of the intervals between successive
		views, and then we estimate the parameters of power law distribution fit
		to these averages. By doing this, if the time that elapsed since we last saw a given video
		and the time we are predicting its popularity for is large, the probability it is requested, is 
		small according to the inter-arrival distribution, thus preventing staleness. 
		We fit to a power law distribution because it has been shown
		in many real-life information diffusion networks that the propagation time can be modelled by 
		a power law distribution(\cite{Mitzenmacher, Newman:01, Barabási:99, zipf:wiki, Mihaljev:10, Capocci:12}). We define,
		\begin{center}
			$\Delta(t)$ - The inter-arrival time probability distribution for videos. \\
			$\Delta(t) = \dfrac{\alpha - 1}{t_{min}}\bigg(\dfrac{t}{t_{min}} \bigg)^\alpha$\\
			$\delta(\mathcal{D}_v(1,t,t))$ = average inter-arrival time in cascade $v$ \\ 
			$(\alpha^*, t_{min}^*) = \argmax\limits_{\alpha, t_{min}} \prod\limits_{v \in \mathcal{V}} \Delta(\delta(\mathcal{D}_v(1,t,t)))$\\
			$zipf(v,t,w)$ - Probability of video $v$ in $[t-w\ t)$ according to the zipf distribution derived from the baseline approach
		\end{center}
		We first find the maximizing $\alpha^*$ and $t_{min}^*$ based on our training
		set $\mathcal{T}$. The score is then calculated as the probability 
		that the video is watched at the time, $t$. The score is given by,
		\begin{equation}
		\label{eq:S-inter}
		S_{\text{inter}}(v,t,w) = \dfrac{zipf(v,t,w) \cdot \Delta(t-l_v(1,t,w))}{\sum\limits_{v'} \bigg[ zipf(v',t,w) \cdot \Delta(t-l_{v'}(1,t,w)) \bigg]}
		\end{equation}
		Using the zipf probability distribution to represent frequency (we use zipf because it is a good approximation
		of the distribution of objects in networks of diffusion \cite{zipf:wiki, Yu:06}) and the inter-arrival distribution to represent
		recency, we give a different angle to analyzing recency and frequency with the notion of staleness which is
		an achilles heel of the LRFU approach.	The ordering for this approach is a sort in decreasing probability.
	\end{subsection}

	\begin{subsection}{Social Approach}
		\label{social_sect}
		This approach is different from the other ones in that it considers users when predicting the
		popularity of a video. It predicts for each user, the probability of that user watching each video.
		And given those probabilities, we estimate the number of views for each video by
		summing over all the users, the probability each user watches that video. Before we can calculate
		the user-video probabilities, we must first estimate the user-user sharing/transmission
		probabilities. These probabilities are modelled as the edges of a diffusion graph between the users of
		the network.\\
		\underline{\bf{Diffusion Model.}}
		In classical epidemiology, we can classify for a given virus $v$, the population into
		(not necessarily disjoint) sets representing the stage each individual is in the life cycle of that virus.
		If they have not yet been infected at time $t$, we say they are in the Susceptible set, $\mathcal{S}(t)$.
		If they have been infected and are infectious, the are in the Infectious set, $\mathcal{I}(t)$ , and if they
		have recovered they go into the Recovered set, $\mathcal{R}(t)$. In this work, we consider the $\mathcal{S}$-$\mathcal{I}$ model,
		where individuals transition from being susceptible to being infectious and remain infectious once infected. The transition
		from $\mathcal{S}$ to $\mathcal{I}$ occurs in two stages, first transmission, then incubation. Before an individual
		can be said to be infected, they must have contracted the virus from a carrier. The difference between the contraction time
		of the infection and the outbreak of symptoms is the incubation time.\\
		For this paper, the set of users is the population of individuals and the videos are the viruses. The probability
		that a user, $u$ gets infected by a video, $v$, at time $t$ is then the same as the probability the individual, $u \in \mathcal{S}(t)$, contracted the 
		infection from an already infected individual, $u' \in \mathcal{I}(t)$, and the incubation time is the difference between $t$ and the time
		that $u'$ transmitted the disease to $u$. In this work we assume that as soon as user gets infected (watches a video), they 
		immediately transmit the video to all other users not yet infected with some probability. Hence the transmission 
		time from $u'$ to $u$ is the infection time, $\tau_{u'}^v$ of $u'$. Let $\chi_v(u',u,t)$ represent the probability that
		user $u'$ infects user $u$ with video $v$ at time $t$.
		\begin{equation}
			\chi_v(u',u,t) = A_{u'u} \cdot \Delta\big(t - \tau^v_{u'}\big)
		\end{equation}
		We learn the transmission probabilities (as an adjacency matrix $A$) under the maximum likelihood framework proposed by Myers et al in \cite{Myers:10},
		and we assume that the incubation times follow a power-law distribution as diffusions in information networks have been shown
		to follow \cite{Mitzenmacher, Newman:01, Barabási:99, zipf:wiki, Mihaljev:10, Capocci:12}. We use the same exponent for the
		power-law distribution as we do for the inter-arrival approach. The sequence/series of infections of a given video is
		called a cascade, and we make an  independent cascade assumption, which means each video is transmitted independent of other videos.\\
		The score for a given video is the sum, across all the users that are not yet infected, of the probability
		that each user is infected at the given time $t$.
		\begin{equation}
		S_{\text{diffusion}}(v,t) = \sum\limits_{u \in \mathcal{S}(t)} 
											\bigg[ 1 - \prod\limits_{u'; \in \mathcal{I}(t)}
							\big(1 - \chi_v(u',u,t)\big) \bigg] \\
		\end{equation}
		The ordering is a descending sort of the diffusion scores.
	\end{subsection}

	\begin{subsection}{Combined Approach}
		\label{approach:combine}
		One deficiency of the social approach is that if there is not enough data to create a complete graph
		based on diffusion, then we are only predicting the views for a small subset of the users 
		in the	network, which will lead to underperformance. A remedy for this is for those users that are not
		part of the diffusion graph, but part of the network, we estimate their probabilities from a consensus
		approach as previously described.\\
		Yet another observation about the social approach is that not every video watched by users that appears
		in the diffusion graph is necessarily fully explained through diffusion, (independent) personal tastes,
		influence from external sources like news sites and blogs, and so on are bound to play roles in affecting
		what the user watches as well. We do not attempt to fully model this phenomenon in this work, but we leave it up 
		to a future work.\\
		The score for the combined approach is given by,
		\begin{center}
			$\tilde{\mathcal{U}}(v)$ - Set of users that have not watched $v$ and are not in the diffusion graph\\
		\end{center}
		\begin{equation}
			S_{\text{combined}}(v,t) = S_{\text{diffusion}}(v,t) + |\tilde{\mathcal{U}}(v)| \cdot S_{\text{inter}}(v,t)
		\end{equation}
		And our rank is given by a descending sort of the combined scores.
	\end{subsection}

\end{section}

\begin{section}{Numerical Analysis and Results}
	\label{numec_analysis}
	Our data is from the University of Massachussetts, Amherst, YouTube network 
	traces\footnote{\url{http://skuld.cs.umass.edu/traces/network/youtube_traces.tgz}} described and
	analyzed by \cite{Zink:09}. We utilize 120 consecutive (from Thu 03/13/2008 19:00 to Tue 03/18/2008 18:10) 
	hours of YouTube requests from their campus network and we partition this data into a training set over the 
	first sixty hours. The training set contains a total of $79213$ requests made by $7260$ unique users over $58345$
	unique videos. The testing set contains a total of $96568$ requests made by $6383$ unique users over $72528$ unique
	videos. In the dataset, there are  total of $10349$ unique users and $120973$ unique videos.
	For our experiments, we make our caching periods units of length, $1$ hour. For each of our these periods, we create a cache $\hat{X}(t,k)$
	and as our performance metric, we look at the average hit rate over all the time periods in our testing set,
	for different values of $k$. \\
	\underline{\bf{Baseline.}}
	As explained earlier, we rank each video in decreasing order according to its approximate popularity under the LRFU scheme
	it got during that time. We employ a window size of $w=28$ for our baseline, because empirically on
	our training set we	see that this window size gives the best average hit rate as shown in \figurename\ref{window_size}. \\
	\begin{figure}
	\centering
	\includegraphics[width=2.5in]{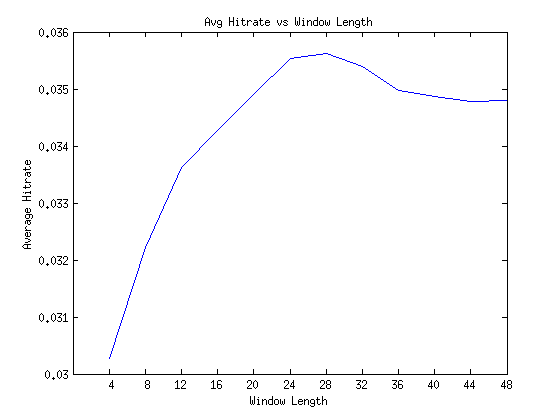}
	\caption{The effect of window size on average hit rate. Window size 28 gives the best improvement.}
	\label{window_size}
	\end{figure}
	\begin{figure}
	\centering
	\includegraphics[width=3.5in]{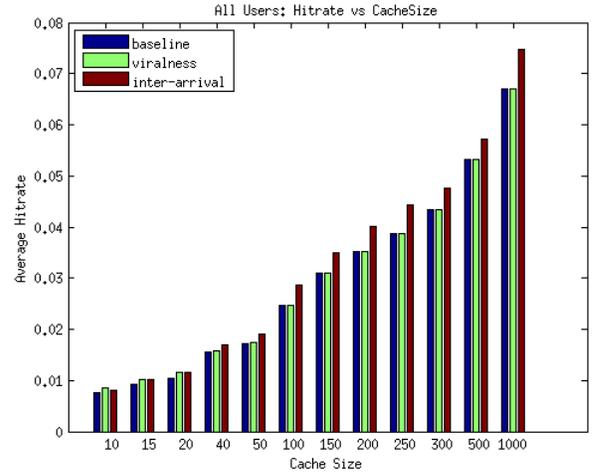}
	\caption{The comparison between the consensus approaches. For the smaller cache sizes, the viralness approach performs better than the other
					consensus approaches, and as discussed in section \ref{approach:baseline}, we see that the viralness approach converges to the baseline.
					Over all cache sizes, we can see that the inter-arrival approach demonstrates significant improvements over the baseline.}
	\label{consensus_fig}
	\end{figure}
	\underline{\bf{Viralness.}}
	For the viralness approach, we have a constraint on the lower bound for the number of requests
	in useful cascades from our testing set. For any video to be considered for the cache, $\hat{X}(t,k)$, it must
	have at least five views($|\mathcal{D}_v(1,t,w)| \geq 5$).	We also add another constraint that these requests are made 
	over at least three time	periods (hours in our case), so we can sufficiently examine the growth trend over time, i.e.,
	$l_v(1,t,w) - l_v(0,t,w) \geq 3$. Let the number of videos that fit these criteria be $n$. If $k > n$, we populate the
	remaining $k-n$ entries of the cache according to the order that the remaining videos occur in the baseline.
	For a given network trace, $n$ is fixed so as $k$ increases, the baseline and the viralness approach should converge.\\
	We compare the results of this approach ($S_{viral}(v,.3,.8,1,t,w)$) with the baseline (\figurename\ref{consensus_fig}),
	and we notice that for a small cache size ($k \leq 50$), we are performing better than the baseline, and after that 
	the results converge as per intuition. On the small cache sizes, we get an improvement of about
	6.1\% over baseline by using the viralness approach.\\
	\underline{\bf{Inter-arrival Time.}}
	For this approach we also have the constraint that the cascades used in learning the parameters
	for the inter-arrival distribution must be of at least length five, i.e., $|\mathcal{D}_v| \geq 5$. This is so
	the averages of the inter-arrival times for the videos we will eventually be fitting to are less noisy.\\
	We then calculate the scores as described in section \ref{p_state}, using the output of our baseline method as the 
	input for the zipfian distribution \cite{zipf:wiki, Yu:06} used in eq.(\ref{eq:S-inter}). We compare the result of this approach to 
	our baseline (\figurename\ref{consensus_fig}) and we get on average about a  11.5\% improvement in hit rate.\\
	\underline{\bf{Social.}}
	To learn the incoming transmission probabilities on the training set, we must also learn power-law distribution parameters.
	We learn these parameters exactly as in the inter-arrival time approach. Another constraint in choosing valid cascades
	to learn the incoming transmission probabilities from is, $|\mathcal{U}(\mathcal{D}_v)| \geq 3$, where 
	$\mathcal{U}(\mathcal{D}_v) = \{u: (u,v,t) \in \mathcal{D}_v\}$. That is, not only must
	the cascade be at least 3 requests long, but the cascades must also have at least 3 unique users. This is an attempt
	to remove noise by increasing the probability that one of those views was as a result of sharing between users.\\
	On our testing set, we relearn the adjacency matrix every ten (10) hours, and limit ourselves to only inferring from the past $w=60$ 
	hours of history each time we learn a new adjacency matrix. We use the algorithm described by Myers et al. in \cite{Myers:10} with 
	a sparsity of 300.\\
	After these parameters and transmission probabilities are learned, we proceed to calculate the future scores of the video for each of the 
	periods in our testing set. For each hour, we consider as prediction history all the requests that were made in the last $w=16$
	hours. We calculate the score using the function described in section \ref{social_sect}, with the appropriate adjacency matrix.\\
	\begin{figure}
	\centering
	\includegraphics[width=3.5in]{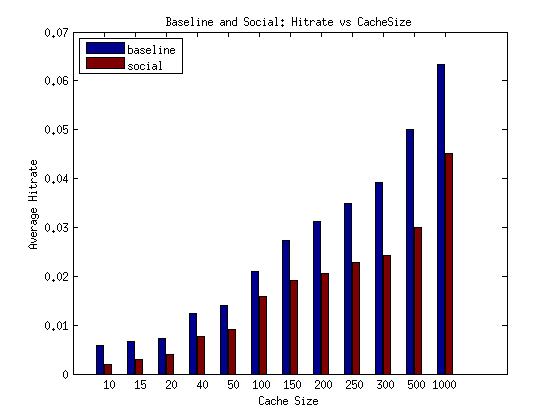}
	\caption{The comparison between the Social approach and Baseline on all users. 
					As was mentioned in section \ref{approach:combine}, the social approach
					suffers from the fact that only 40\% of all users have non-zero request probabilities.}
	\label{base_vs_social}
	\end{figure}	
	Because of the constraints on valid cascades, we find that on average only about $40\%$ of the users in the network are used
	for inference, which implies only $40\%$ of the nodes in the graph are in a connected component and the others
	are just isolated nodes. This ultimately leads to under-performance (\figurename\ref{base_vs_social}) since for about $60\%$ of the users we can
	make no estimation of the probability it views a given video. So in order to guage the usefulness of this
	approach, we also  run it on an augmented dataset (\figurename\ref{connected_base_vs_social}) where only users in the connected component are in the 
	dataset ($\tilde{\mathcal{T}} = \bigcup\limits_{\mathcal{D}_v; |\mathcal{U}(\mathcal{D}_v)| \geq 3} \mathcal{D}_v$).
	\begin{figure}
	\centering
	\includegraphics[width=3.5in]{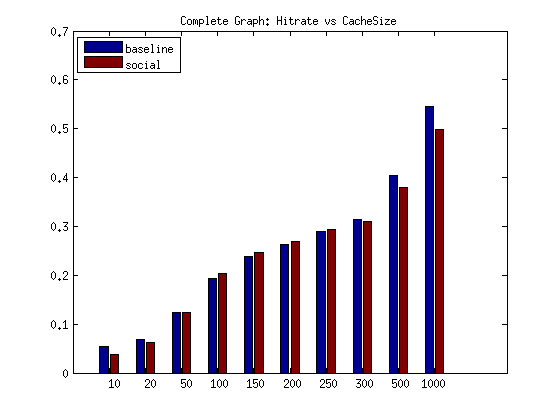}
	\caption{The comparison between the Social approach and Baseline when on connected users. We can see from this that the social popularity distribution
					which implicitly takes into account recency and frequency through diffusion is within 5\% of the approximate popularity distribution based on LRFU
					if indeed all the users are governed by diffusion}
	\label{connected_base_vs_social}
	\end{figure}\\
	\underline{\bf{Combined.}}
	As previously noted in section \ref{approach:combine}, it is unlikely that even in a 
	connected graph, the volume of videos watched will be completely accounted for 
	by diffusions over the graph. In \cite{Myers:12}, Myers et al., showed that only about
	71\% of information volume on twitter could be accounted for by network diffusions.
	To that end, on our data we performed an experiment to figure the best weighting (ranging from 0\% to 100\% in steps of 10)
	between the social scores and the inter-arrival scores, and, corroborating the conclusion of \cite{Myers:12}, it came out to
	be 70\% from social and 30\% from inter-arrival.\\
	We also  analyze the performance of the combined approach under the 
	full data, and under just the connected component. Under the full dataset, we see an improvement
	of 13.2\% from the baseline to the combined (\figurename\ref{full_combined}),	and under just
	the connected components, our improvement rises to 21.1\% (\figurename\ref{connected_combined}).\\
	\begin{figure}[h]
	\centering
	\includegraphics[width=3.5in]{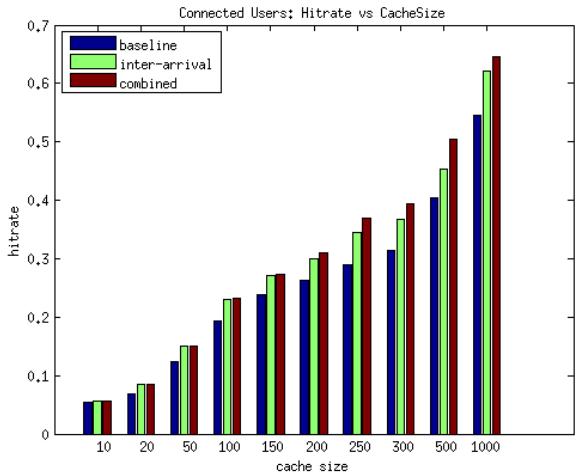}
	\caption{The comparison between the Combined and Inter-Arrival on connected users. Starting from medium size
					cached, the combined approach outperforms the inter-arrival approach.}
	\label{connected_combined}
	\end{figure}
	\begin{figure}[h]
	\centering
	\includegraphics[width=3.5in]{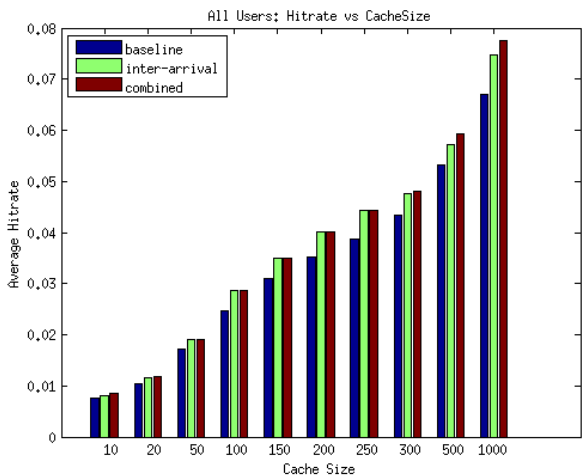}
	\caption{The comparison between the Combined and Inter-Arrival on all users. The combined approach outperforms the
					inter-arrival approach especially on the larger cache sizes, but still caches more relevant videos even on smaller
					cache sizes.}
	\label{full_combined}
	\end{figure}
	 We also compare the performance of the combined to the inter-arrival approach. From
	our experiments we see that the combination of social and inter-arrival gives
	an improvement of 1.6\% over inter-arrival on this full dataset (\figurename\ref{full_combined}),
	and 4.8\% when only users from the connected component are used(\figurename\ref{connected_combined}).\\
	\ \\
\begin{table}[!t]
\renewcommand{\arraystretch}{1.3}
\caption{Summary of Results on All Users}
\label{all_results}
\centering
\begin{tabular}{|c|c|c|}
\hline
\bfseries Method A & \bfseries Method B& \bfseries \% improvement\\
\hline
Baseline & Viralness (small cache) & 6.2 \\
\hline
Baseline & Inter-Arrival & 11.6\\
\hline
Baseline & Combined & 13.2\\
\hline
Inter-Arrival & Combined & 1.6\\
\hline
\end{tabular}
\end{table}
	From our results, we see that for the social approach to give considerable gain, it 
	is imperative that the underlying structure be a connected graph, and not a
	graph with mostly isolated vertices. We have also seen that the combination
	of the social approach with a consensus approach (inter-arrival) generally
	outperforms any of the individual approaches. This is because on this dataset,
	and as we suspect on most social networks, neither diffusion, nor consensus can
	fully explain the request patterns of the users in the network. Users tend to have
	their own preferences, and are also influenced by different external media like
	the news, or blogs, and such behavior has been studied by Myers et al. in \cite{Myers:12},
	where they show that only about 71\% of information volume on twitter can be
	attributed to network diffusion.

\begin{table}[!t]
\renewcommand{\arraystretch}{1.3}
\caption{Summary of Results on Connected Users}
\label{connected_results}
\centering
\begin{tabular}{|c|c|c|}
\hline
\bfseries Method A & \bfseries Method B& \bfseries \% improvement\\
\hline
Baseline & Inter-Arrival & 15.6\\
\hline
Baseline & Combined & 21.1\\
\hline
Inter-Arrival & Combined & 4.9\\
\hline
\end{tabular}
\end{table}

\end{section}

\begin{section}{Robustness and Complexity}
  Our algorithms described in section \ref{p_state} have some free parameters
  that are application and data specific. Although, for our application, we did not
  exhaustively search the parameter space for the optimal choice of these
  parameters, we chose our parameters based on what we believe to be reasonable
  assumptions. For example, our choices of what cascades to use to perform the 
  inference was made under the assumption that cascades of shorter lengths
  could lead to over-fitting our model to what might be noise, hence we choose
  only cascades of some minimum length (elaborated on in section \ref{numec_analysis}).
  We leave the analysis of the performance of our algorithms under varying
  (exhaustively) parameter settings up for future work. \\
  On the run-time of the social approach, as noted by Myers et al. in \cite{Myers:10},
  the columns of the adjacency matrix can be inferred independently of each other
  which leads to opportunity for massive parallelization. In our experiments, each 
  column took on average 3 seconds to be inferred solving the optimization
  problem via the KNITRO optimization software in MATLAB \cite{knitro:matlab}.
  This means that for our network of about 4000 nodes, it takes about 16
  minutes running 12 jobs simultaneously. And since we update our matrices only
  every 10 hours, this seems a very reasonable cost to us. Given the adjacency matrix,
  computing the scores for our diffusion approach takes $O(|\mathcal{V}|\cdot|\mathcal{U}| + 
  |\mathcal{V}|\log|\mathcal{V}|)$ time, versus $O(\mathcal{V}|\log|\mathcal{V}|)$  for the
  baseline. The first term in the complexity of the diffusion approach comes from
  computing for every user the probability that they watch every video (which is the 
  expected amount of each video transmitted to each user). The logarithmic term 
  comes from the sorting of the scores. For our networks, $|\mathcal{V}| \gg 
  |\mathcal{U}|$, hence the complexity of the baseline and our approach are
  comparable.
  
\end{section}

\begin{section}{Conclusions}
	In this paper we have shown that by leveraging social cues inferred from the requests
	made by users in the network, through an estimated latent social graph over these users,
	we can better predict the popularity distribution of the videos being requested by the users. 
	Our preferred method of combining the distributions resulting from the social-approach and
	the inter-arrival (staleness) approach is shown to outperform other approaches. This is because our model
	captures the idea of videos being spread like diseases over a network of users where some users
	are more likely to infect (and be infected by) other users, which the other approaches do not. These
	considerations result in a 14\% improvement over the baseline.
\end{section}

\begin{section}{Future Work}
	One of the roadblocks we faced in this work is the inadequate amount of data both in terms of
	recency and volume, so an immediate follow up to this work is to gather more recent
	data on a longer scale from different network sites and verify our findings from this work.
	We also aim to address the issue of the robustness of the algorithms to varying of the parameters,
	and the performance when the uniform video size assumption is lifted.\\
	Also in terms of future directions on different approaches, we will like to explore
	the application of large alphabet prediction, preferential attachment graphical models,
	and possibly predictive sparse coding on the this problem. \\
\end{section}

\begin{section}{Acknowledgements}
	\label{shout-out}
	This work is in part supported by Intel-Cisco-Verizon via the VAWN program.
\end{section}

\bibliographystyle{IEEEtran}
\bibliography{myBib}

\end{document}